\documentclass[doublecol]{epl2}

\usepackage{color}
\usepackage{amsmath}
\usepackage{amssymb}
\usepackage{graphicx}
\usepackage{bm}
\usepackage{hyperref}

\def\be{\begin{equation}}
\def\ee{\end{equation}}

\author{Pierre-Philippe Cr\'epin \inst{1,2} \and Xavier Leyronas  \inst{3} \and Fr\'ed\'eric Chevy  \inst{1} }
\shortauthor{Cr\'epin \etal}

\institute{
  \inst{1} Laboratoire Kastler Brossel, ENS-PSL Research University, CNRS, UPMC, Coll\`ege de France, 24, rue Lhomond, 75005 Paris\\
  \inst{2} D\'epartement de physique, \'Ecole Polytechnique, 91128 Palaiseau cedex, France\\
  \inst{3}  Laboratoire de Physique Statistique, \'Ecole Normale
Sup\'erieure, PSL Research University, Universit\'e Paris Diderot Sorbonne
Paris-Cit\'e, Sorbonne Universit\'es UPMC Univ Paris 06, CNRS, 24 rue Lhomond,
75005 Paris, France
}

\title{Hydrodynamic spectrum of a superfluid in an elongated trap}


\date{today}
\abstract{In this article we study the hydrodynamic spectrum of a superfluid confined in a cylindrical trap. We show that the dispersion relation $\omega(q)$ of the phonon branch scales like $\sqrt{q}$ at large $q$, leading to a vanishingly small superfluid critical velocity. In practice the critical velocity is set by the breakdown of the hydrodynamic approximation. For a broad class of superfluids, this entails a reduction of the critical velocity by a factor $(\hbar\omega_\perp/\mu_c)^{1/3}$ with respect to the free-space prediction (here $\omega_\perp$ is the trapping frequency and $\mu_{\rm c}$ the chemical potential of the cloud).}

\begin{document}

\maketitle

\section{Introduction}

Superfluidity is one of the most spectacular manifestations of quantum mechanics at the macroscopic scale and has been under intense theoretical and experimental study since the beginning of the XXth century. One of the key features of superfluids is the existence of hydrodynamic modes leading to dramatic effects such as first and second sounds \cite{atkins1959liquid}. In a seminal work, L. Landau related the dispersion relation of the elementary excitations of a superfluid to the critical velocity above which a quantum liquid cannot sustain a frictionless flow \cite{landau41theory}. This critical velocity has been observed in Liquid Helium and ultracold gases as well \cite{raman1999evidence, Chikkatur2000Suppression, miller2007, desbuquois2012superfluid, weimer2014critical,ferrier2014mixture} but over  the past few years, experiments using dilute Bose-Einstein condensates (BEC), fermionic vapours, and superfluid Bose-Fermi mixtures, have raised questions about the applicability of Landau's argument in trapped systems.  Ref. \cite{fedichev2001critical,tozzo2003bogoliubov,steinhauer2003bragg,singh2015probing} showed that for an elongated weakly interacting Bose-Einstein condensate, the transverse trapping was responsible for a significant reduction of the critical velocity, due to a back-bending of the dispersion relation when the wave-length becomes comparable to the transverse size of the cloud \cite{zaremba1998sound,stringari1998dynamics}. In this article, we extend this work to an arbitrary quantum liquid and we focus on the long wavelength, hydrodynamic excitations of the system. Using a combination of analytical calculations and scaling arguments, we show that the sub-linearity of the dispersion relation at short wave-length is a robust result that  depends neither on the nature nor the equation of state of the superfluid and that for a broad class of physical systems, it leads to a reduction of the critical velocity by a factor $\sim (\mu_c/\hbar\omega_\perp)^{1/3}$, where $\mu_c$ is the chemical potential of the cloud and $\omega_\perp$ the transverse trapping frequency.

\section{General scaling arguments}

Consider a superfluid described by a zero-temperature equation of state $\mu=f(n)$ relating the chemical potential $\mu$ to the density $n$. In the following, we will assume that $f$ can be written as $f(n)=n^\gamma g(n)$ where $g(n)$ has a well-defined limit $C$ for $n$ going to 0. For a Bose-Einstein condensate, $\gamma=1$ while for a fermionic superfluid, we have $\gamma=2/3$ for $a<0$ and $\gamma=1$ for $a>0$ (for the unitary case $a=\infty$, we have $\gamma=2/3$ and $g$ is actually a constant) \cite{zwerger2012BCSBEC}. When the wavelength of the excitations of the system is larger than the coherence length $\xi$ of the superfluid, its dynamics  can be described within the so-called hydrodynamic formalism where  density $n$ and velocity $\bm v$ are solutions of \cite{stringaripitaevskii}
\begin{eqnarray}
m\left(\partial_t \bm v+\bm \nabla v^2/2\right)&=&-\bm\nabla\left(V(\bm r)+\mu (n)\right)\\
\partial_t n+\bm \nabla (n \bm v)&=&0,
\end{eqnarray}
with $V$ the trapping potential. For a static system ($\bm v=0$), the density profile $n_0$ is given by the Local Density Approximation (LDA) equation
\be
V(\bm r)+\mu(n_0(\bm r))=\mu_{\rm c},
\ee
where $\mu_{\rm c}$ is the global chemical potential of the cloud.
For weak excitation, we can write $n=n_0+n_1$, with $n_1\ll n_0$ and after expanding the hydrodynamic equations to first order in perturbation, we see that $\mu_1=\partial_{n_0}\mu_0n_1$ is solution of \cite{stringari1998dynamics}
\be
\partial_t^2\mu_1-\frac{\partial\mu_0}{\partial n_0}\bm\nabla\left(\frac{n_0}{m}\bm\nabla\mu_1\right)=0,
\ee
where $\mu_0(\bm r)=\mu(n_0(\bm r))$.

 We consider here an elongated harmonic trap described by a potential energy $V(\bm r)=m\omega_\perp^2(x^2+y^2)/2$. Owing to the space and time-translational symmetries, we can look for exponentially varying solutions

 $$
 \mu_1(\bm r,t)=\tilde\mu_1(x,y)e^{i(qz-\omega t)},
 $$
where $\tilde\mu_1$ satisfies

\be
-\omega^2\tilde\mu_1-\frac{\partial\mu_0}{\partial n}\bm\nabla_\perp\left(\frac{n_0}{m}\bm\nabla_\perp\tilde\mu_1\right)+q^2\frac{n_0}{m}\frac{\partial\mu_0}{\partial n}\tilde\mu_1=0.
\label{EqHydro}
\ee

Let us first focus on the low-energy sector of the dispersion relation. In this limit we expect the existence of a phonon branch described by a sound velocity \cite{stringari1998dynamics,Capuzzi2006Sound}
\be
c_{\rm 1D}^2=\frac{\bar n_0}{m}\frac{\partial \mu}{\partial\bar n_0},
\ee
where the 1D density $\bar n_0$ is defined as
\be
\bar n_0=\int d^2\bm\rho n_0(\bm r),
\ee
with $\bm\rho=(x,y)$ the projection of the position in the transverse plane. Taking $\mu\sim\mu_c$ and $\bar n_0\sim n_0(0) R_\perp^2$, where $R_\perp$ is the transverse size of the cloud, we therefore expect the sound velocity to scale like $\sqrt{\mu_c/m}$. Since according to the Local Density Approximation (LDA), $\mu_{\rm c}=m\omega_\perp^2R_\perp^2/2$, we have $c_{\rm 1D}\sim \omega_\perp R_\perp$.

\begin{figure}
\centerline{\includegraphics[width=\columnwidth]{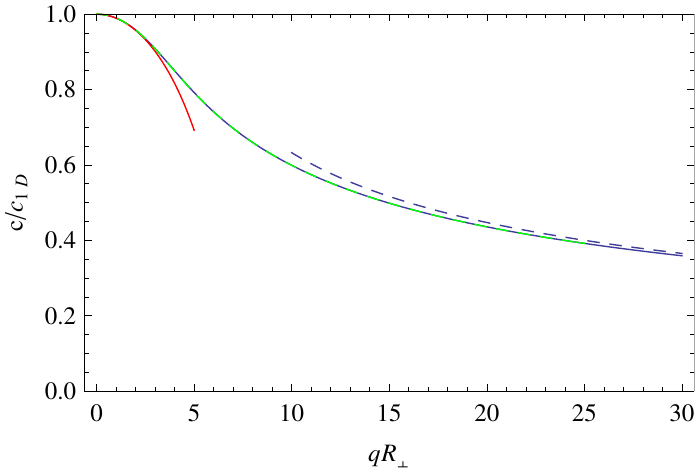}}
\caption{Phase velocity of a trapped superfluid (here a mean-field BEC). Red solid line: low momentum expansion (Eq. (\ref{Eq:LowQ})); Dashed blue line: high momentum expansion (Eq. \ref{ScalingOmega}); The exact numerical resolution (solid blue) and the variational calculation (dashed green, Eq. \ref{Eq:Variationnel}) cannot be distinguished on this scale.}
\label{Fig2}
\end{figure}

In the large-momentum limit, Eq. (\ref{EqHydro}) predicts a sublinear behaviour of the dispersion relation, as first pointed out in \cite{stringari1998dynamics,zaremba1998sound} for a weakly interacting BEC and we attribute this property to the localization of the high frequency modes in the wings of the cloud where the density, hence the sound velocity, is lower. Indeed,  the last term of Eq. (\ref{EqHydro}) dominates over the others  for large $q$ unless $\tilde\mu_1$ takes significant values only in regions where $n_0\partial_n\mu_0$ is small, i.e. in the wings of the cloud. Assume then that the perturbation $\tilde\mu_1$ is localized within a distance $\lambda\ll R_\perp$ from the cloud edge. In this low density region we can approximate the equation of state by its dilute limit $\mu=C n^\gamma$ and since $\mu_0=\mu_c(1-\rho^2/R_\perp^2)$, we can approximate $\mu_0$ by $\mu_0\sim \mu_c \lambda/R_\perp$. We then have the following scalings

\begin{eqnarray}
\frac{\partial\mu_0}{\partial n}\bm\nabla_\perp\left(\frac{n_0}{m}\bm\nabla_\perp\tilde\mu_1\right)&\sim& \frac{\mu_c\tilde\mu_1}{m\lambda R_\perp}\\
q^2\frac{n_0}{m}\frac{\partial\mu_0}{\partial n}\tilde\mu_1&\sim& q^2\mu_c\frac{\lambda}{m R_\perp}\tilde\mu_1.
\end{eqnarray}
Equating the three terms of Eq. (\ref{EqHydro}) we expect that
\be
\omega^2\sim \frac{\mu_c}{m\lambda R_\perp}\sim  q^2\mu_c\frac{\lambda}{m R_\perp}
\ee
yielding the following scaling
\begin{eqnarray}
\lambda&\sim&\frac{1}{q}\label{ScalingLambda}\\
\omega&\sim&\omega_\perp\sqrt{R_\perp q}.\label{ScalingOmega}
\end{eqnarray}
Eq. (\ref{ScalingLambda}) confirms that the short-wavelength modes are localized on the edges of the cloud while Eq. (\ref{ScalingOmega}) demonstrates  the sublinear behavior of the dispersion relation at large momenta. We summarize the different scalings in Fig. \ref{Fig2}, where we compare them to an exact numerical resolution of the hydrodynamic equations.

This $\sqrt{q}$ behavior has a dramatic consequence on the superfluid critical velocity. Indeed, according to Landau's original argument, frictionless flows cannot exist for velocities larger than
\be
v_{\rm c}=\min_q \left(\frac{\omega(q)}{q}\right),
\ee
and consequently, the large-momentum behavior of the dispersion relation predicts a vanishingly small critical velocity in a harmonic trap.

In a real system, the scaling (\ref{ScalingOmega}) is not valid for arbitrarily large $q$, but only as long as $q\xi\ll 1$ \footnote{Note that the restriction does not invalidate Eq. (\ref{ScalingLambda}) and (\ref{ScalingOmega}) that are valid as long as the hydrodynamical hypothesis $q\xi\ll 1$ is satisfied as well as the short wavelength condition $\lambda\ll R_\perp$. Since $\lambda\sim 1/q$, we see that the validity domain of the $\sqrt{q}$ behaviour is finite as long as $R_\perp\gg\xi$.}. Hydrodynamics will thus breakdown for momenta larger than a typical value $q^*$ defined by
\be
q^*\sim\frac{1}{\xi (\mu_c/q^*R_\perp)},
\ee
where we evaluate $\xi(\mu)$ at the typical chemical potential $\mu_c\lambda/R_\perp\sim \mu_c/q^*R_\perp$.  If we assume that $v_c\sim c(q^*)$, we finally obtain
\be
v_{\rm c}\sim\frac{\omega_\perp R_\perp}{\sqrt{q^* R_\perp}}\sim \frac{c_{\rm 1D}}{\sqrt{q^* R_\perp}}\\
\ee

\begin{figure}
\centerline{\includegraphics[width=\columnwidth]{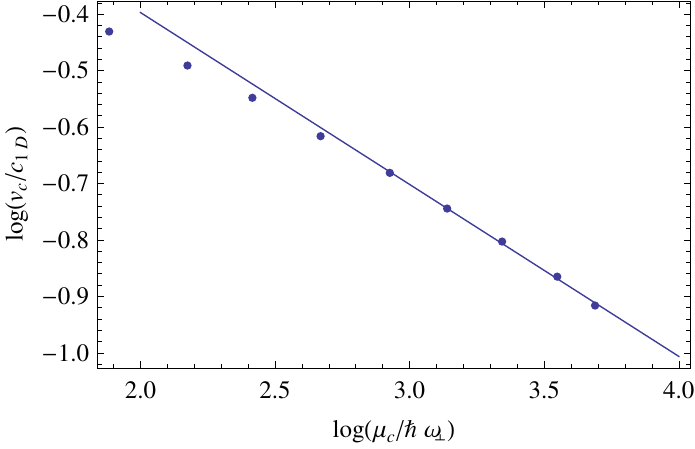}}
\caption{Comparison between scaling Eq. \ref{Eq:vc} for the superfluid critical velocity of a weakly interacting Bose-Einstein condensate and the numerical results obtained from direct resolution of the Gross-Pitaevskii Equation \cite{fedichev2001critical} (points). The dashed line corresponds to a fit of the last four points by a power law $v_c=A c_{1D}(\hbar\omega_\perp/\mu_{\rm c})^{1/3}$, where $A$ is the only free parameter. A fit to the results of \cite{fedichev2001critical} yields $A\simeq 1.2$ for a mean-field Bose-Einstein condensate.}
\label{Fig1}
\end{figure}

For a weakly interacting BEC or a unitary Fermi gas, we have $\mu\sim\hbar^2/m\xi^2$ leading to the simple scaling
\be
v_{\rm c}\sim \omega_\perp R_\perp \left(\frac{\hbar\omega_\perp}{\mu_c}\right)^{1/3}\sim c_{\rm 1D} \left(\frac{\hbar\omega_\perp}{\mu_c}\right)^{1/3},
\label{Eq:vc}
\ee
where we have used the fact that $c_{\rm 1D}\sim\sqrt{\mu_c/m}\sim \omega_\perp R_\perp$. A similar reduction of the critical velocity was found numerically in \cite{fedichev2001critical} for a weakly interacting Bose-Einstein condensate and we see in Fig. \ref{Fig1} that the $(\hbar\omega_\perp/\mu_c)^{1/3}$ suppression factor agrees with these numerical results.

\section{Low-$q$ behaviour}

We now analyze more quantitatively the solutions of Eq. (\ref{EqHydro}) and we focus first on the large-wavelength limit. We rescale the different parameters of the problem by taking $\mu'=\mu/\mu_c$, $n'=n/n_c$ and $\bm\rho'=\bm\rho /R_\perp$. In these dimensionless units Eq. (\ref{EqHydro}) becomes

\be
-\frac{\omega^2 m}{q^2\mu_c}\tilde\mu_1+\frac{1}{q^2R_\perp^2}\frac{\partial\mu'_0}{\partial n'}\bm\nabla'_\perp\left[n'_0\bm\nabla'_\perp \tilde\mu'_1\right]+n_0'\frac{\partial\mu'_0}{\partial n'}\tilde\mu'_1=0.
\label{EqDimensionless}
\ee

Take $\varepsilon=\omega^2 m/q^2\mu_c$. Eq. (\ref{EqDimensionless}) shows that $\varepsilon$ is a function of the dimensionless parameter $q'=qR_\perp$ and we compute it for small $q$ using standard perturbation techniques. Since we focus here on the phonon branch, we have $\omega\simeq c_{\rm 1D} q$ for small $q$, hence $\varepsilon_{q'\rightarrow 0}\simeq c_{\rm 1D}^2m/\mu_c$. We therefore write

\begin{eqnarray}
\varepsilon&=&\sum_{k=0}^{\infty}\varepsilon_k q'^k\\
\tilde\mu_1&=&\sum_{k=0}^{\infty}\tilde\mu_1'^{(k)} q'^k.
\end{eqnarray}
To leading order, the perturbative expansion yields
\be
\frac{\partial\mu'_0}{\partial n'}\bm\nabla'_\perp\left[n'_0\bm\nabla'_\perp \tilde\mu'^{(0)}_1\right]=0
\ee
hence $\tilde\mu'^{(0)}_1={\rm cte}$.

To next order, we have then
\be
-\varepsilon_0\tilde\mu'^{(0)}_1+\frac{\partial\mu'_0}{\partial n'}\bm\nabla'_\perp\left[n'_0\bm\nabla'_\perp \tilde\mu'^{(1)}_1\right]+n_0'\frac{\partial\mu'_0}{\partial n'}\tilde\mu'^{(0)}_1=0.
\ee
Multiply this equation by $\partial_{\mu'} n'_0$ and integrate over transverse coordinates. The middle term then vanishes and we are left with
\be
-\varepsilon_0\int d^2\bm \rho'\frac{\partial n'_0}{\partial\mu'}+\int d^2\bm\rho' n_0'=0.
\ee
Using the fact that within LDA we have $d\mu=m\omega_\perp^2 \rho d\rho$ as well as Gibbs-Duhem relation $dP=nd\mu$ we finally obtain
\be
\varepsilon_0=\frac{P_0(\mu_{\rm c})}{\mu_c n_0(\mu_{\rm c})},
\ee
 hence
 \be
 c_{\rm 1D}=\sqrt{\frac{P_0(\mu_{\rm c})}{mn_0(\mu_{\rm c})}}.
 \ee
In the case of a polytropic equation of state  $\mu=C n^\gamma$, this expression yields back the known result
\be
c_{\rm 1D}=\sqrt{\frac{\gamma\mu_{\rm c}}{m(\gamma+1)}}
 \ee
 derived previously is \cite{Capuzzi2006Sound}.

Pushing the expansion one step further, we obtain
\be
\frac{\omega^2}{q^2}=c_{\rm 1D}^2(1-\alpha q^2R_\perp^2),
\label{Eq:LowQ}
\ee
with
\be
\alpha=\frac{c_{\rm 1D}^2}{n_0(0)\omega_\perp^2R_\perp^2}\int_0^{R_\perp}\frac{d\rho}{n_0(\rho)\rho}\left[n_0(\rho)-n_0(0)\frac{P_0(\rho)}{P_0(0)}\right]^2
\ee
For a purely polytropic equation of state, the integral can be computed explicitly and we have in this case
\be
\alpha=\frac{\gamma^3 }{4(1+\gamma)^2(1+2\gamma)}.
\label{Eq:polytropic}
\ee
For $\gamma=1$ (mean-field BEC) we get $\alpha=1/48$, as found in \cite{stringari1998dynamics}, while for a unitary Fermi gas we obtain $\alpha=2/175$.

For an arbitrary equation of state we observe that $\alpha$ is positive, meaning that the dispersion relation is always (locally) concave, leading to a critical superfluid velocity smaller than $c_{\rm 1D}$.

\section{Large-$q$ behaviour}
To clarify the origin of the sub-linearity of the dispersion relation, we now turn to the opposite limit of short-wavelength modes. As discussed in the introduction, we expect the perturbation to be localized in the wings of the cloud. In this region, we can replace the equation of state by its low-density expansion $\mu=C n^\gamma$. Taking $x=q(R_\perp-\rho)$ in Eq. (\ref{EqHydro}), we have to leading order in $x$
\be
x\gamma\tilde\mu_1=\gamma x\frac{d^2\tilde\mu_1}{dx^2}+\frac{d\tilde\mu_1}{dx}+\frac{\omega^2}{\omega_\perp^2qR_\perp}\tilde\mu_1.
\label{EqSchrodinger}
\ee
We note that this equation is an eignevalue problem parametrized by $\beta=\omega^2/\omega_\perp^2qR_\perp$. Assuming that the value of $\beta$ is known, we see that the dispersion relation is
\be
\omega=\omega_\perp\sqrt{\beta q R_\perp},
\ee
where we recover the $\sqrt{q}$ scaling derived in the first part of the article.

To find $\beta$, we note that Eq. (\ref{EqSchrodinger})  is similar to Shr\"odinger's equation for the hydrogen atom and can be solved in a similar way. We find that the lowest excitation branch is characterized by a chemical potential
\be
\tilde\mu_1(\rho)=e^{q(\rho-R_\perp)}
\label{Eq:Schrodinger}
\ee
yielding $\beta=1$ hence the expected scaling
\be
\omega = \omega_\perp\sqrt{qR_\perp}.
\ee

\section{Variational approach and interpolation}
\begin{figure}
\centerline{\includegraphics[width=\columnwidth]{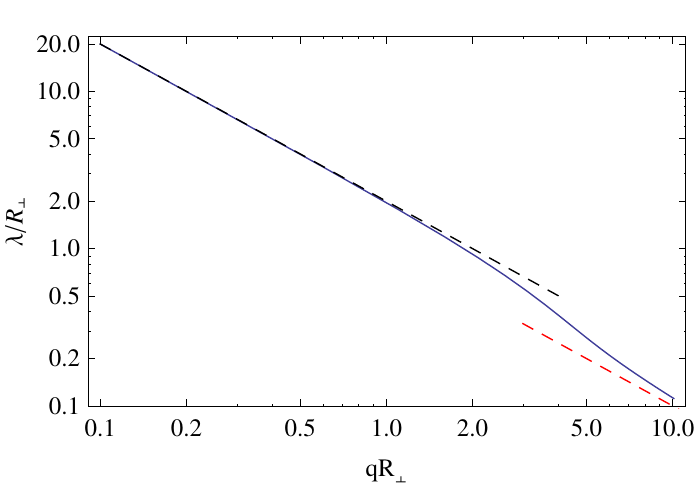}}
\caption{Solid line: Variational value of $\lambda$ for a BEC; Red dashed line: high momentum expansion $\lambda\simeq 1/q$. Black dashed line: low-momentum asymptotic behavior $\lambda\simeq 2/q$, in agreement with Eq. (\ref{Eq:variationelLowQ}) for $\gamma=1$.}
\label{Fig3}
\end{figure}

To support the previous analytical results, we provide a variational resolution of the problem interpolating between the two regimes. For this purpose, we first note that Eq. (\ref{EqHydro}) can be formally written as
\be
\omega^2\tilde\mu_1={\cal L}_{\bm q}[\tilde\mu_1],
\ee
where the linear operator ${\cal L}_{\bm q}$ is defined by
\be
{\cal L}_{\bm q}[u]=-\frac{\partial\mu_0}{\partial n}\bm\nabla_\perp\left(\frac{n_0}{m}\bm\nabla_\perp u\right)+q^2\frac{n_0}{m}\frac{\partial\mu_0}{\partial n}u.
\ee
${\cal L}_{\bm q}$ is a positive and hermitian operator for the scalar product
\be
\langle u|v\rangle=\int d^2\bm\rho\left(\frac{\partial n}{\partial\mu}\right)_0u^*(\bm\rho)v(\bm\rho).
\ee
Using a standard variational scheme, we can look for an approximate value of $\omega(\bm q)$ by minimizing

\be
\tilde\omega^2=\frac{\langle u|{\cal L}_{q}[u]\rangle}{\langle u|u\rangle}
\label{Eq:Variationnel}
\ee
over a well-chosen functional space. From the asymptotic analysis performed at small and large $q$'s we know that in the long-wavelength limit $\tilde\mu_1$ is uniform while it increases exponentially close to $\rho=R_\perp$   at short-wavelength. A possible variational Ansatz satisfying also the mirror symmetry across the $z$ axis is $u(\bm\rho)=\cosh(\rho/\lambda)$, where $\lambda$ is the variational parameter. By construction, this Ansatz provides the exact result in the two limits studied previously and should in addition offer a reasonable semi-analytical interpolation for arbitrary wavelength perturbation. We represent in Fig. \ref{Fig3} the value $\lambda$ for a Bose-Einstein condensate that confirms the two asymptotic regimes discussed above. For small $q$, $\lambda$ diverges, corresponding to a uniform perturbation in the transverse plane. More precisely, by taking $u\simeq 1+\rho^2/2\lambda^2$, we find the analytic expression for $\lambda$

\be
\lambda=\frac{1}{q}\sqrt{2\frac{\partial\ln \bar n_0}{\partial\ln \langle\rho^2\rangle}},
\ee
with $\langle\rho^2\rangle=\bar n_0^{-1}\int d^2\bm\rho n_0\rho^2$. For a polytropic equation of state, this expression simplifies into
\be
\lambda=\frac{1}{q}\sqrt{\frac{2(1+\gamma)}{\gamma}}.
\label{Eq:variationelLowQ}
\ee
Note that when inserting this expression for $\lambda$ into the variational expression Eq. (\ref{Eq:Variationnel}), we recover Eq. (\ref{Eq:polytropic}) for the large wavelength expansion of the dispersion relation.  In the opposite large $q$ limit, the width of the perturbed region scales as $\lambda\simeq 1/q$ as obtained already in Eq. (\ref{Eq:Schrodinger}). The corresponding velocity is displayed in Fig. \ref{Fig2} where we see that it agrees almost perfectly with the exact numerical resolution of the hydrodynamic equations.

\section{Conclusion}
In this article we have calculated the hydrodynamic spectrum of a superfluid confined in an elongated harmonic potential. We showed that the short wavelength modes are localized in the wings of the cloud and follow a sub-linear dispersion relation $\omega_q\simeq \omega_\perp \sqrt{q R_\perp}$, this behaviour being valid in the range $1\ll qR_\perp\ll R_\perp/\xi$. This behaviour leads to a strong suppression of the critical velocity below which the superfluid can sustain a frictionless flow. These predictions are well within reach of current experiments and can be tested both bosonic and fermionic superfluids in the BECè-BCS crossover. They may then explain the suppression of the critical velocity observed in some experiments.

\acknowledgments

The authors acknowledge support from R\'egion Ile de France (Atomix Project), ERC (ThermoDynaMix Project) and Institut de France (Louis D. Prize). They thank the ultracold Fermi group for fruitful discussions.

\bibliographystyle{unsrt}
\bibliography{bibliographie}

\end{document}